# A microfluidic platform for high-throughput multiplexed protein quantitation.

Francesca Volpetti, Jose Garcia-Cordero†, and Sebastian J. Maerkl[*]

Institute of Bioengineering, School of Engineering, Ecole Polytechnique Federale de Lausanne, Lausanne, Switzerland.

† Current address: CINVESTAV-Monterrey, PIIT, Nuevo León 66600, Mexico.

E-mail: sebastian.maerkl@epfl.ch

**Abstract**

We present a high-throughput microfluidic platform capable of quantitating up to 384 biomarkers in 4 distinct samples by immunoassay. The microfluidic device contains 384 unit cells, which can be individually programmed with pairs of capture and detection antibody. Samples are quantitated in each unit cell by four independent MITOMI detection areas, allowing four samples to be analyzed in parallel for a total of 1,536 assays per device. We show that the device can be pre-assembled and stored for weeks at elevated temperature and we performed proof-of-concept experiments simultaneously quantitating IL-6, IL-1β, TNF-α, PSA, and GFP. Finally, we show that the platform can be used to identify functional antibody combinations by screening 64 antibody combinations requiring up to 384 unique assays per device.

**Introduction**

The precise quantitation of proteins is important in systems biology and is becoming of interest in clinical studies. In both cases it is increasingly necessary to monitor dozens if not hundreds of proteins per sample, to provide an overview of protein levels in signalling networks [1], or to derive higher order correlations in clinical samples [2,3]. Techniques for multiplexed analysis of proteins have therefore attracted considerable attention [4].

The classical and still widespread approach for quantitating proteins is based on immunoassays, particularly ELISA, which provides high-specificity, sensitivity and dynamic range, but is relatively low-throughput and extremely cost-ineffective due to large reagent and sample consumption. To alleviate some of these problems low to medium throughput microfluidic methods have been developed for protein quantitation [5–7]. Multiplexed alternatives to ELISAs have recently been developed including array based detection schemes [2,8] and bead based detection [9–11]. These approaches enabled the parallel analysis of large numbers of biomarkers. One significant difference between multiplexed bead assays and standard ELISAs is the fact that all antibody pairs are allowed to cross-react in the multiplexed bead assay, giving rise to potential off-target effects. A recent refinement of an array based approach circumvented these limitations using a method called antibody co-localization microarray (ACM) which specifically co-localizes antibody pairs using a microarrayer and has been used to analyse up to 50 proteins in parallel [12]. Although the ACM approach solved the problem of cross-reactivity, showed low limits of detection (LODs) and a high dynamic range, it remains difficult to implement. Microwestern arrays (MWA) were developed to alleviate some of the various shortcomings of bead-based multiplexed analysis, reverse-phase lysate arrays, and mass spectrometry based approaches [13]. MWAs are miniaturized and partially automated western blots. With MWAs it is possible to analyse 192 proteins in 6 samples for a total of 1,152 measurements. Nonetheless, MWAs are difficult to fully automate as all processing steps required for a standard western blot such as electrophoretic separation, transfer to a nitrocellulose membrane, and development with antibodies are still necessary and the entire process requires 14-24 hours.

Microfluidic based approaches are appealing alternatives to these existing methods, as all fluid handling steps can be integrated and thus automated. A recent microfluidic device based on mechanically induced trapping of molecular interactions (MITOMI) [14] could analyse 4 biomarkers in 8 samples, or vice versa [6]. We have recently reported highly-integrated and parallelized microfluidic nanoimmunoassays based on MITOMI coupled to sample microarrays, allowing the analysis of 4 biomarkers in 384 samples per device [15] and more recently 4 biomarkers in 1,024 human clinical serum samples, for a total of 4,096 measurements, requiring as little as 5 nL per sample [16]. Here we demonstrate a new functional mode enabling the parallel analysis of up to 384 biomarkers in 4 samples, for a total of 1,536 measurements per device. In our approach, each antibody pair is isolated in its own unit cell, completely eliminating cross-talk. We show that we can achieve good sensitivity and high-dynamic range with MITOMI. The approach requires ~5-25 nl (0.45-2.25 ng) of antibody per assay and is thus extremely cost effective. Antibody suppliers generally provide 100 mg of antibody, which is sufficient for ~100,000 assays using our approach. We also demonstrate that the device and antibody array can be assembled and consequently stored for weeks in ambient conditions prior to use. All processing steps are fully integrated on the microfluidic device and can be automated [17]. We measured IL-6, IL-1β, TNF-α, PSA, and GFP, in buffer and determined the limit of detection (LOD, > mean + 2 std. deviations of the control) for these 5 proteins to be 4, 4, 30, 15, and 4 pM, respectively. We go on to show that the platform can be applied to identify optimal antibody combinations of a 3 antibody immunosandwich assay by screening all 64 combinations of 4 primary biotinylated anti-PSA mouse monoclonal antibodies, 4 secondary anti-PSA rabbit poly- and monoclonal antibodies, and 4 tertiary anti-rabbit IgG fluorophore conjugated antibodies. We tested multiple titrations of the secondary antibody for each of the antibody combinations to derive precise, quantitative information on the relative binding strength for all 64-antibody combinations. We found that many antibody combinations were incompatible and thus non-functional, but nonetheless were able to identify several antibody combinations that performed well. The platform should thus find applications were it is necessary to perform multiplexed biomarker analysis or in combinatoric screens.

**Experimental**

**Device fabrication**.
The device is composed of two PDMS layers (control and flow layer) aligned to an epoxy-functionalized microscope slide. The devices were fabricated by multilayer-soft-lithography [18]. SU8 GM1060 photoresist was used to pattern the 30 μm high control layer, which was spin coated first for 10 s at 500 rpm, then for 20.3 s at 2527 rpm, once more at 2527 rpm for 40 s, then for 1 s at 3527 rpm, followed by 1 s at 2527 rpm. After a soft bake step the wafer was exposed on a MA6 mask aligner for 8 s followed by a post exposure bake. The exposed wafer was developed in PGMEA, followed by a hard bake at 160°C. AZ9260 photoresist was used for the 10 μm high flow layer. It was spin coated for 10 s at 800 rpm, followed by 40 s at 1800 rpm. After a soft bake the wafer was exposed on an MA6 mask aligner for two intervals of 18 s. The flow layer mold was annealed at 180°C for 30 min to round the flow channels. 20:1 and 5:1 ratios of elastomer (Sylgard 184) were used for the flow and the control layer, respectively. These were cured at 80°C for 30 min and then aligned. After aligning the flow and control layer were bonded at 80°C for 1.5 h.

**Epoxy-silane coated glass slide**
The protocol consists of two parts: cleaning of the glass slides followed by coating them with epoxy silane. In the first step a solution of milli-Q water and ammonia solution ($NH_4OH$ 25%) in a 5:1 ratio was heated to 80°C. When the solution reached said temperature, 150 mL of hydrogen peroxide ($H_2O_2$ 30%) was added to the bath. The glass slides were immersed in this solution for 30 min. Glass slides were then rinsed with milli-Q water and dried with nitrogen. In the second step the slides were submerged in a 1% solution of 3-Glycidoxypropryl-trimethoxymethylsilane (97% pure) in toluene for 20 min. The slides are then rinsed with fresh toluene to remove any unbounded 3-GPS, dried with nitrogen and followed by a baking step at 120°C for 30 min. The glass slides were sonicated in fresh toluene for 20 min and afterwards rinsed with isopropanol and dried. The epoxy-coated slides were stored in a vacuum chamber at room temperature until used.

**Microarray spotting**
A high-throughput microarray platform (Genetix QArray2) was used to array the reagents. Depending on the experiment a combination of primary, secondary, tertiary antibody, and sample were spotted. A 4.9 nL delivery-volume spotting pin (946MP8XB, Arrayit, USA) was used to array all samples. Each spot was spotted 5 times to increase reagent concentrations. After each sample the pin was washed 3 times using $dH_2O$ for 500 ms, between each washing step the pin was dried for 500 ms. The humidity in the spotting robot was set to 60%.

**Antibodies and proteins**
Biotinylated goat polyclonal antibody to GFP was purchased from Abcam (ab6658), EGFP from Biovision (4999-100) and the anti Penta-His Alexa Fluor 647 conjugate was purchased from Qiagen (35370). A matched monoclonal anti human-PSA antibody pair and purified native human PSA were purchased from Fitzgerald Industries International (MA, USA) (10-P20E, 10-P20D, 30C-CP1017U). Anti-PSA antibodies were purchased from Abcam (ab53774, ab76113, ab19554) and from Fitzgerald (70R-35481). Anti-PSA biotin conjugated antibodies were purchased from Abcam (ab182031, ab77310, ab182030) and from Abcore (1B-289). The anti-cytokine antibodies were purchased from eBioscience: IL-6 biotin-conjugate, IL-1β biotin-conjugate, TNF-α, (13-7068-81, 13-7016-81, 16-7384-85). The anti human-IL-6 (12-7069-81), IL-1β (13-7016-81), TNF-α (12-7349-81) labeled with phycoerythrin (PE), human IL-6 protein standard (39-8069-65), human IL-1β protein standard (39-8018-65), human TNF-α protein standard (39-8329-65) were purchased from eBioscience. Goat anti-Rabbit IgG labelled with phycoerythrin (Pe) were purchased from Abcam (ab72465, ab97070) and from Life-Technologies (P2771MP), Goat anti-Rabbit IgG labelled with Alexa Fluor 546 from Life-Technologies (A-11010). Neutravidin biotin-binding protein and the biotinylated bovine serum albumin BSA were purchased from Thermo Scientific (29130 and 31000 respectively).

Primary antibodies for PSA (10-P20D) and TNF-α were biotinylated using a ChromaLink One-Shot Antibody Biotinylation kit (Solulink) according to the manufacture instructions. Anti human PSA (10-P20E) was labeled with PE using the R-PE Antibody All-in-One Conjugation kit (Solulink) according to the supplied protocol.

**Device operation**
Microfluidic control channels were primed with $dH_2O$ (or oil: Fluorinert FC-40) starting at 5 psi, once the channels were completely filled, the pressure was increased to 15 psi to close the fluidic valves and to 20 psi to actuate the MITOMI button membranes (Figure 1.a (3)). All reagents were aspirated into Tygon tubing (i.d. 0.51 mm, o.d. 1.52 mm) and connected to the flow channels, and the flow-channel pressure was set at 3.5 psi [15].

**Data analysis**
The microfluidic device was scanned using a fluorescent microarray scanner (ArrayWorx e-Biochip Reader, Applied Precision, USA). The images were exported as 16-bit TIFF files and analyzed using microarray image analysis software (GenePix Pro v6.0, Molecular device). Non-linear regression analysis of the data was performed using Prism 5.0 (GraphPad).

**Results and discussion**

**Assay workflow**
The device is fabricated by multilayer soft-lithography [18] and is comprised of two PDMS layers (control and flow layer) aligned to an epoxy functionalized glass slide pre-arrayed with antibody pairs (Figure 1.a). A DNA microarrayer was used to array the pairs of primary and secondary antibodies, which subsequently are aligned to the unit cells so that a microfluidic chamber encloses each antibody spot. The aligned device is bonded to the epoxy-coated glass slide by incubating overnight at 40°C (Figure 1.b).

The microfluidic design is composed of 384 unit cells [15]. Each unit cell consists of two 1.7 nl antibody chambers, one for the primary antibody and the other for the secondary antibody. A detection chamber is located between the two antibody storage chambers and includes 4 detection regions created by MITOMI buttons (120 μm diameter). Each unit cell can be isolated with a microfluidic valve, and the antibody chambers are separated from the reaction chamber when the sample is flown through the main channel (Figure 1.a).

MITOMI is a micro-mechanical method recently developed to allow the quantitative analysis of molecular interactions [14,19]. MITOMI consists of a freestanding "button" membrane, which can be actuated by pneumatic or hydraulic pressure, similarly to standard micro-mechanical valves generated by multilayer soft-lithography [18]. When the button membrane is actuated it physically contacts a circular area on the glass surface of the microfluidic device. When the button membrane is in contact with the glass surface it protects the surface from solute and solvent molecules and can thus be used for surface patterning [20]. In addition to surface patterning, MITOMI is primarily used to mechanically trap surface bound molecules between the derivatized glass surface and the PDMS button membrane preventing dissociation of these molecules and thus allowing the precise measurement of transient molecular interactions. MITOMI has previously been applied to characterizing protein-DNA [14,21–24], protein-RNA [25,26], protein-protein [27], protein – small molecule interactions [26] and kinetic measurements [23,27].

The assay itself is composed of four steps: i) functionalization of the surface, ii) surface immobilization of the primary antibody, iii) flow of the sample, iv) incubation with the secondary antibody, and readout (Figure 1.c). To functionalize the surface, a solution of 2 mg/ml biotin-BSA PBS is flown through the detection chambers for 20 min with the buttons open, followed by a wash with 0.005% Tween PBS for 10 min. Next 0.5 mg/ml neutravidin PBS was flowed for 20 min, followed again by a wash step. After this step the buttons were actuated in order to specifically protect the detection areas while the remaining surface was passivated with biotin-BSA, which was flowed for 20 min followed by a wash step.

Next, the unit cells are isolated and the biotinylated primary antibody allowed to diffuse into the reaction chamber. The buttons were kept closed until the antibody equilibrated to allow a homogeneous immobilization of the primary antibody. After 2.5 h the buttons were opened for 30 min and the biotinylated antibodies were immobilized to the four neutravidin coated detection regions. We then closed the MITOMI buttons to protect the detection regions from cross-contamination, opened the sandwich valves that previously isolated the unit cells, closed the primary antibody chamber, and washed the detection region with 0.005% Tween PBS. Four different samples can be measured on the device as each of the four MITOMI buttons can be independently actuated. It is thus possible to either measure one sample at 4 different dilutions, which increases the dynamic range of the assay, or 4 independent samples. Each sample is flowed for 20 min, followed by a 10 min wash step. The MITOMI buttons are actuated during the wash step to prevent dissociation and thus loss of antigen, and cross-contamination. Finally, the unit cells are once again isolated, and the secondary antibody allowed to diffuse into the detection area for 2.5 h. Each button is then opened sequentially for 20 min to allow association of the secondary antibody. The sequential opening at this step further reduces any possibility of cross-contamination between buttons, which is particularly important if independent samples are tested. The entire device is then scanned with a standard DNA microarray scanner, and antigen is detected and quantitated via the fluorescently labeled secondary antibody.

**Assay optimization**
The concentrations of the primary and secondary antibody were optimized in order to achieve optimal device performance and low limits of detection. We spotted 10 different concentrations of primary antibody to maximize signal while minimizing the required antibody amounts. After surface functionalization, the primary antibodies were re-hydrated and allowed to diffuse into the reaction chamber. We then flowed 30 nM of GFP over all four buttons in the open state for 20 minutes. GFP pull-down saturated at a spotted antibody concentration of 600 nM (Figure 2.a), and we consequently spotted all primary antibodies at a concentration of 600 nM. Special care was taken to ensure that all four buttons were uniformly derivatized with primary antibody, which required keeping the buttons closed while the primary antibody was allowed to equilibrate for ~2 hours. Failure to do so gave rise to higher antibody densities on the two buttons near the antibody source chamber, which could not be remedied with longer incubation times, indicating that the amount of spotted antibody is not significantly larger than the amount that can be surface immobilized.

We next optimized the secondary antibody concentration. Different dilutions of anti-GFP secondary antibodies labeled with Alexa Fluor 647 and GFP were spotted on the glass slide to simultaneously assess the impact of antigen and secondary antibody concentration. For this experiment we simply flowed the primary antibody for 20 min at a concentration of 30 nM. The fluorescence signal from the antibody-antigen-antibody complex (Figure 2.b) began to saturate when the concentration of the secondary antibodies reached a value of 28 nM. The limit of detection (LOD) is ~10 pM (> mean + 2 std. deviations of the control) and a dynamic range of 3 orders of magnitude can be achieved in the range of 1.2-140 nM for the secondary antibody. Since the LOD was not significantly different over a secondary antibody range of 1.2 – 140 nM we opted to use a concentration of 6 nM, which resulted in low levels of non-specific signal, a low LOD, and a good dynamic range. Higher secondary antibody concentrations give rise to higher signal but didn't improve the LOD. Furthermore, the use of lower antibody concentrations can result in cost savings when a very large number of assays are to be performed. But since our platform already uses very small quantities of antibody per assay these considerations are secondary as a single vial of antibody (100 μg) is sufficient for ~100,000 assays. Nonetheless, higher secondary antibody concentrations do perform better when high biomarker levels are to be quantitated, and our platform does allow for the parallel use of multiple antibody concentrations in different microfluidic unit cells, which is appealing if high-dynamic range samples are to be quantitated. For the following multiplex experiments we chose 600 nM and 6 nM for the primary antibody and the secondary antibody concentrations, respectively.

**Multiplex assay**
To demonstrate the multiplexing capability of the platform, we tested various combinations between 5 antibody pairs on a single device for 1,408 experiments (Figure S1). GFP primary antibody was tested against all 5 secondary antibodies each at two different concentrations, and a BSA negative control, for a total of 11 combinations. These combinations were spotted 4 times requiring 44 unit cells. The remaining 4 primary antibodies (IL-6, IL-1β, TNF-α, PSA) were tested against 5 different secondary antibodies each at two concentrations and one BSA control, but spotted in 7 replicates, requiring a total of 308 unit cells. In total, 352 unit cells on the device were used in this experiment. We measured a negative control and an antigen cocktail containing 5 proteins in buffer (IL-6, IL-1β, TNF-α, PSA, GFP) each at two concentrations and one BSA control, but spotted in 7 replicates giving rise to 1,408 experiments. The solutions were introduced on-chip starting with the BSA control and then from the lowest to the highest concentration of the antigen cocktail, one for each button. As shown in Figure 3 (Figure S2) the specific signal from each reaction is considerably higher than the non-specific specific signal, indicating that, at least for these antibody combinations, no considerable cross-reactivity was observed.

Depending on the antigen the LOD varies from 4 pM to 30 pM. In our previous work [16] LODs of 3.67 pM, 742 fM, 897 fM and 1.04 pM were achieved for TNF-α, IL-6, IL-1β and PSA, respectively. These LODs are comparable to other microfludic methods. Zheng et al. report a MITOMI device capable of detecting 10-100 pg/ml (55.6 – 556 fM) of CEA [6], Fan et al. report a sensitivity of <1-30 pM for various cytokines using a barcode based microfluidic platform [2], and Pla-Roca et al. measured LODs for TNF-α = 15.6 pg/ml (917 fM), Il-6 = 42.9 pg/ml (2.04 pM), and Il-1β = 12.3 pg/ml (683 fM) [12]. Commercially available macroscale assays often exhibit higher sensitivities. Meso Scale Discovery reports the following sensitivities: TNF-α = 1.1 pg/ml (64.7 fM), Il-6 = 0.7 pg/ml (33.3 fM), and Il-1β = 2.4 pg/ml (133.3 fM), and Luminex: TNF-α = 25 pg/ml (1.47 pM), Il-6 = 5 pg/ml (238.1 fM), and Il-1β = 15 pg/ml (833.3 fM).

**Combinatoric screen to identify functional antibody combinations**
Immunoassays generally require two or three antibodies, two of which directly bind to the biomarker of interest, while the third is required for readout. It can thus be non-trivial to identify combinations of antibodies that function in concert, particularly the two antibodies that are required to bind to the same protein biomarker. In addition to this combinatoric problem of identifying antibodies that work well together, it is also of interest to identify those antibodies that have the highest affinity for their antigen. Our multiplexed immunoassay platform is ideally suited to conduct large-scale screens to address these issues.

To test whether our platform can be used to identify functional antibody combinations we chose to optimize a 3-antibody immunoassay for PSA. We acquired 4 primary biotinylated anti-PSA monoclonal antibodies, 4 secondary anti-PSA antibodies, and 4 tertiary anti-rabbit IgG antibodies labeled either with phycoerythrin or Alexa-Fluor 546 (see Table S1), originating from 4 different antibody suppliers. In addition to testing all 64 antibody combinations we also decided to titrate the secondary antibody, to derive relative binding affinities of these antibodies.

We performed the combinatoric screen in 3 different ways. In the first experiment we spotted the 4 secondary antibodies, each at 6 different concentrations (including a no-antibody control), and the 4 tertiary antibodies generating 96 unique combinations (4x4x6) in quadruplicate requiring all 384 unit cells on the device. We then surface immobilized the 4 biotinylated primary antibodies on the 4 MITOMI detection areas to generate 384 unique conditions performed in quadruplicate for a total of 1,536 assays. We then tested the antibody combinations for their ability to detect 100pM of PSA (Figure S3). This experiment provided a global overview of the performance of the various antibody combinations, and showed that two of the secondary antibodies (#1 and #2) were non-functional, or performed poorly. We next repeated the experiment with spotted primary and secondary antibodies and testing 3 of the 4 tertiary antibodies (Figure S4). We opted to remove the δ tertiary antibody from this screen as it gave consistently high non-specific fluorescent background. This second screen returned qualitatively the same results as the first screen. We then repeated the screen on 4 separate devices by spotting 12 concentrations of the secondary antibody, immobilizing one of the 4 biotinylated antibodies in all four MITOMI detection areas, and testing each fluorescently labeled tertiary antibody on a separate MITOMI button for a total of 768 unique conditions (12x4x4x4) with 8 repeats gathering a total of 6,144 data points (Figure 4).

We found no major differences in performance between the 4 primary antibodies tested, but observed a considerable variation in the secondary and tertiary antibody performances. The secondary polyclonal antibody #1 from Fitzgerald repeatedly failed in all combinations tested and the two polyclonal antibodies (#3 and #4) from Abcam consistently outperformed the monoclonal antibody #2 from the same supplier. Both tertiary antibodies (γ and δ) from Abcam gave non-specific signal, apparently because of poor antibody purification after conjugation to phycoerythrin. Nonetheless tertiary antibody δ performed exceedingly well in our screen, but did give rise to considerable non-specific signal when used in combination with primary antibody B. The second best performer of the tertiary antibodies was antibody α from Life-Technologies. These results show that there is a significant variability in antibody performance, but that it is possible to identify functional antibody combinations rapidly, and cost effectively on our multiplexed immunoassay platform.

**Stability test**
We envision these immunoassay devices to ultimately be used either in research or clinical settings, which requires that the assembled microfluidic device and reagents contained therein be stable for extended periods of time, preferably at ambient conditions. We therefore performed a stability test to determine how long the device maintained its functionality. Six slides were spotted with primary and secondary antibodies against GFP. After bonding the PDMS device with the spotted glass slides the assembled devices were stored at 40°C for up to one month. We tested devices periodically after one, two, and four weeks under the same experimental conditions. After one week at 40°C we observed no significant decrease in signal from the stored chip as compared to a new device (Figure 5). After two weeks the signal decreased by ~20% and after one month the signal decrease by about 50%. This study indicates that assembled devices are indeed functional after 1 month stored at elevated temperatures and that after four weeks the decrease in signal is within acceptable limits.

In this series of experiments we assembled the devices, stored them at elevated temperatures, and then performed all fluidic operations. This still requires the use of biotinylated-BSA, and neutravidin. These reagents should be reasonably stable in solution or in lyophilized form, and thus no hindrance to applications in a variety of environments. We nonetheless tested whether it is possible to assemble a device, perform the surface chemistry, dry and store the device. The use of oil (Fluorinert FC-40) in the control layers prevented premature re-hydration of the spotted antibodies. We observed that the low-viscosity oil could slowly seep through the PDMS, leading to fouling of the flow layer and the surface chemistry. Storing the device in vacuum over-night eliminated this problem and allowed us to perform a functional immunoassay one day after generating the surface chemistry and drying the device. A better solution would be to remove the oil from the control lines after generating the surface chemistry, eliminating the need for vacuum storage. The signals obtained from this device were lower than from devices with a freshly prepared surface, but sufficiently high to conduct quantitative measurements (Figure S5). It is thus possible to decouple surface preparation from the actual assay with good results (LOD of 10 pM instead of 4 pM), although the approach could be further optimized (choice of oil, generating a more stable surface chemistry, etc.). Nonetheless, these studies indicate that devices can be pre-assembled and stored, effectively decoupling chip production and utilization.

**Conclusions**

We developed a microfluidic device able to perform multiplexed analysis of up to 384 biomarkers in four samples for a total of 1,536 immunoassays per device. The consumption of antibodies is reduced to nanogram amounts, drastically decreasing assay cost. The primary hurdle to applying this platform to the detection of hundreds of protein biomarkers is the initial acquisition cost of a large number of antibody pairs, which was prohibitive for this proof-of-concept study. Nonetheless the drastically reduced antibody consumption of our platform means that once a library of antibodies is acquired it becomes a long-lasting resource sufficient for ~100,000 assays. Once initial antibody pairs are acquired and validated, they can be used for thousands of assays also eliminating problems associated with antibody batch-to-batch variability, which otherwise would require significant re-validation efforts especially in clinical settings. The platform can of course be used to measure a smaller number of biomarkers. For example, 96 antibody pairs can be spotted in quadruplicate to measure 96 biomarkers in four samples with higher accuracy than could be achieved with a single measurement. We recently demonstrated that each MITOMI button can be used to analyze 3 different samples [20], which when applied to the current device would increase the sample throughput from 4 to 12 samples and 4,608 assays per device.

To validate the platform, four human antigens IL-6, IL-1β, TNF-α, and PSA were detected achieving LODs in the range of 4-30 pM. We also applied the platform to a high-throughput combinatoric screen to identify functional antibody combinations. We showed that the chip can be stored after assembly for extended periods of time in adverse conditions (40°C) enabling the possibility to deploy such devices as academic or clinical research tools. All remaining processing steps are readily automated, and the final method only requires user intervention to introduce a wash solution and samples onto the device. With our current device the antibody derivatization steps require a significant amount of time (2-3 hrs). These steps could be drastically accelerated by including active mixing on the device [28,29]. Active mixing could reduce each step to below 10-15 min, allowing the entire assay to be performed in less than 1-2hrs. These results show that MITOMI based high-throughput protein quantitation is an appealing approach for applications in clinical and basic research. Potential applications of the technology include the analysis of cellular signaling pathways [30] and analysis of clinical samples. A number of clinical investigations are currently underway with the aim to perform full genome and microbiome sequencing coupled to the analysis of a large number of serum proteins [31]. The technology presented here should be an appealing choice for the comprehensive analysis of medium to large panels of protein biomarkers in clinical samples, particularly since we recently demonstrated that a similar microfluidic platform was applicable to measuring protein biomarkers in clinical serum samples [16].

**Figure 1**. **Workflow schematic** a) microfluidic design: the device consists of two PDMS layers: flow (blue) and control (red). The chip is an array of eight rows by 48 columns for 384 unit cells. Each unit cell is composed of: two antibody chambers divided by a reaction chamber (1-2), 4 MITOMI buttons (3), a valve that segregates the unit cells (4), a valve that separates antibody and reaction chambers (5) and a valve for releasing pressure in the antibody chambers (6). b) The PDMS chip is aligned to an epoxy-functionalized slide onto which primary and secondary antibodies were spotted. c) Assay details: schematic of the unit cell and cross section of a button region: i) functionalization of the surface: BSA-biotin is flowed though the chip followed by neutravidin. Next, the buttons are closed and BSA-biotin flowed again to passivate all neutravidin molecules except for those located underneath the MITOMI buttons, ii) the biotinylated primary antibody is allowed to diffuse into the MITOMI detection chamber and is bound by neutravidin immobilizing it in the MITOMI detection regions, iii) the sample is flown through the device and antigens are captured by the surface immobilized antibodies, iv) finally, the fluorescently labeled secondary antibody is allowed to diffuse into the MITOMI area, binds to the antigen if present, and is trapped by MITOMI. The entire device is then quantitated using a DNA microarray scanner.

**Figure 2**. **Optimization of antibody concentration**. a) Different concentrations of anti-GFP primary antibody were spotted and 30 nM of GFP was detected with the four buttons (error bars are std. dev., n=9). All 4 buttons show identical response profiles and 600 nM was chosen as the optimal primary antibody concentration. b) Intensity of fluorescent signal as a function of spotted GFP concentration at 5 different secondary antibody concentrations, the primary antibody was flowed at a concentration of 60 nM. The inset shows the same data on a log-log scale (error bars are std. dev. n=3).

**Figure 3**. **Multiplexed assay and antibody cross-reactivity testing**. Fluorescent signal and binding curves for 5 different proteins (PSA, TNF-α, IL-6, IL-1β, and GFP) quantitated in parallel on a single device. Concentrations for the primary and secondary antibodies used in this experiment were 600 nM and 6 nM, respectively. Each panel shows the signal obtained from the specific primary antibody to the given antigen detected with the various secondary fluorescently labeled detection antibodies. The results show that each antigen was specifically detected by the correct antibody pair and that no significant cross reactivity occurred between the antigens and antibodies tested in this experiment. Left y-axis: phycoerythrin labeled secondary antibody signal; right y-axis: Alexa Fluor 647 labeled secondary antibody signal (error bars are std. dev. n=7).

**Figure 4. Combinatoric antibody screen.** Four different secondary antibodies (1-4) were spotted at 12 different concentrations on four epoxy glass slides and aligned to PDMS chips. For each device one of the four primary antibodies (A-B) was flowed at a concentration of 30 nM through the device and immobilized on the four MITOMI detection regions. Next, PSA was flowed at a concentration of 100 pM. Then, the secondary antibodies were allowed to diffuse into the reaction chamber. Finally, the four different tertiary antibodies (α - δ) were flowed sequentially at a concentration of 30 nM one for each MITOMI button. (error bars are std. dev. n=8)

**Figure 5**. **Stability test**. (left column) Intensity of the fluorescent signal as a function of the GFP concentration flowed through the channel, using a new device and devices stored at 40°C for 1, 2, and 4 weeks. (Right column) Percentage difference between the mean signals measured on the new device and those stored for different amounts of time.

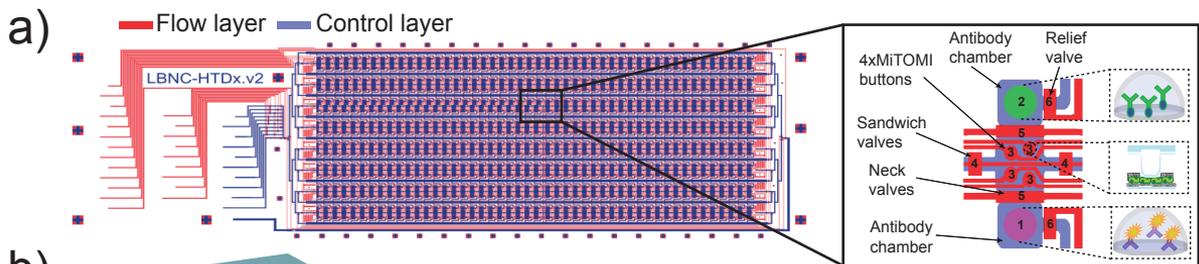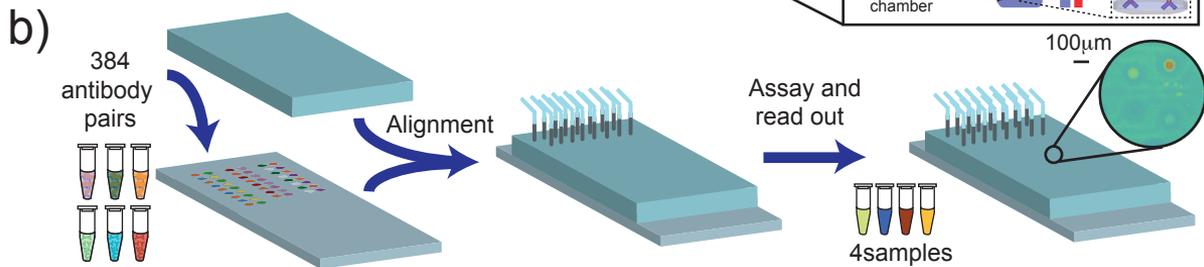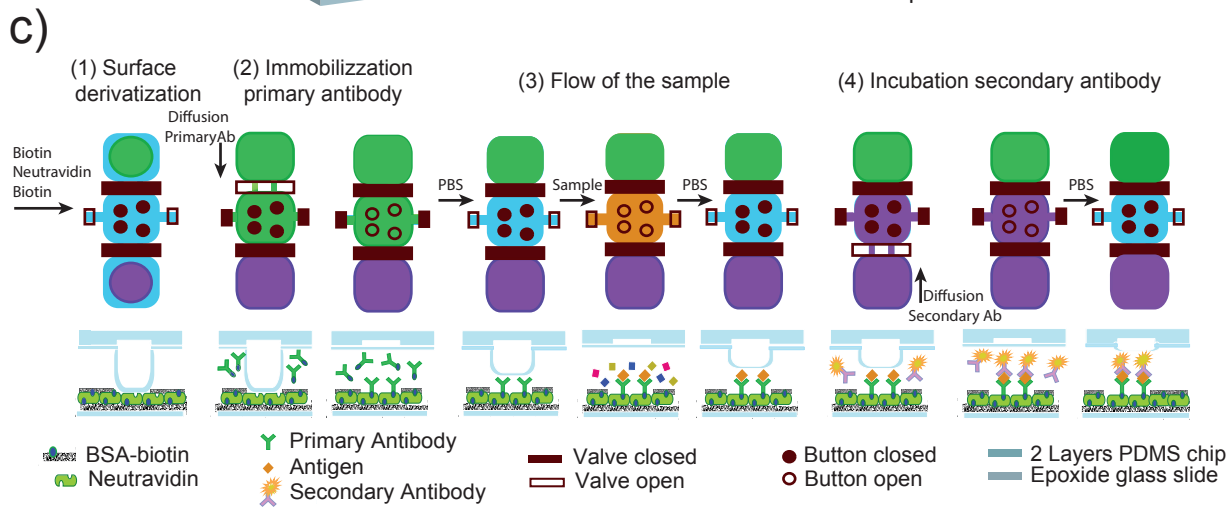

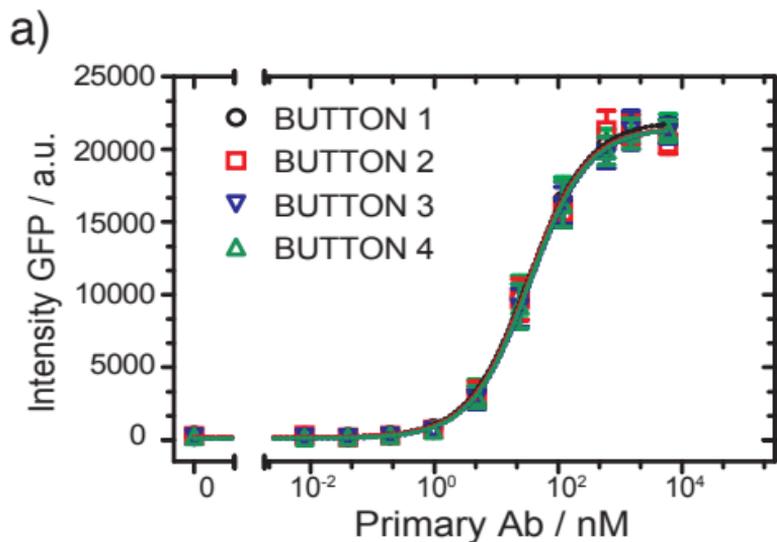
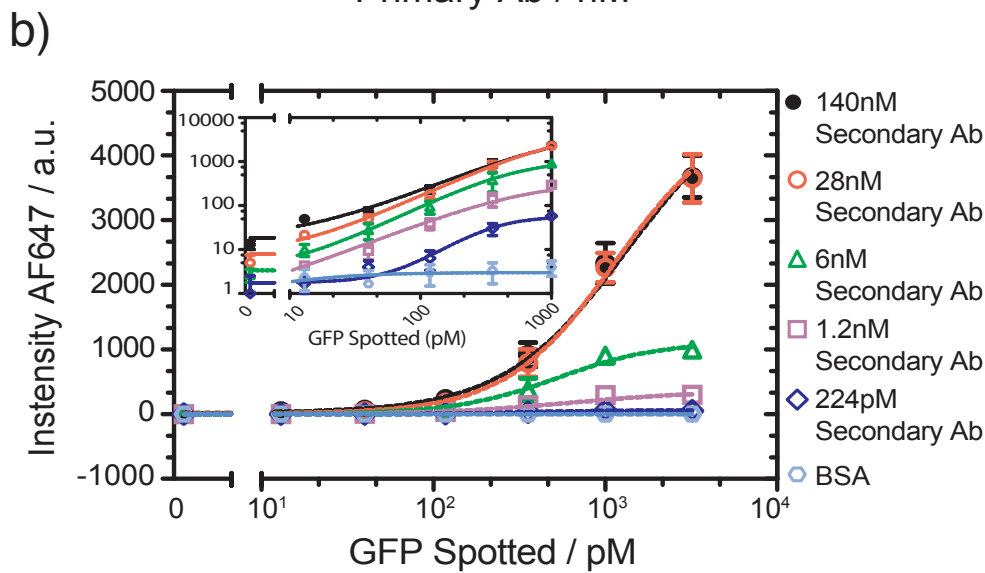

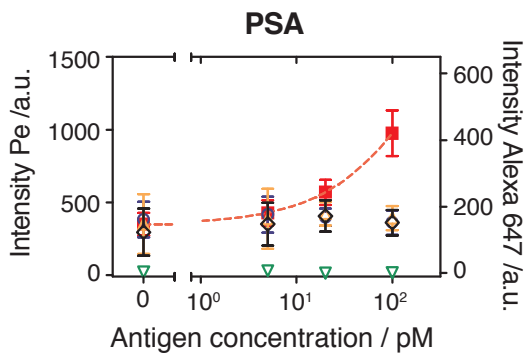
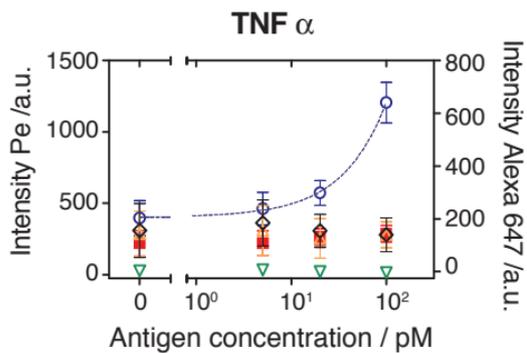
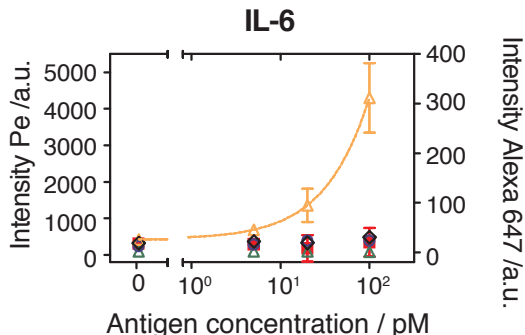
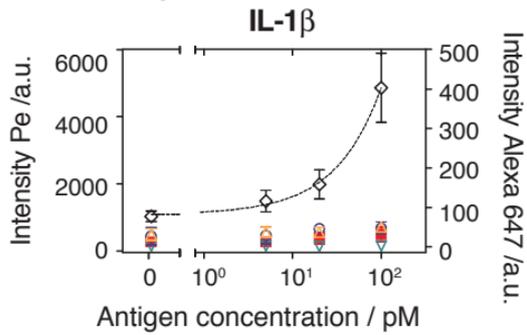
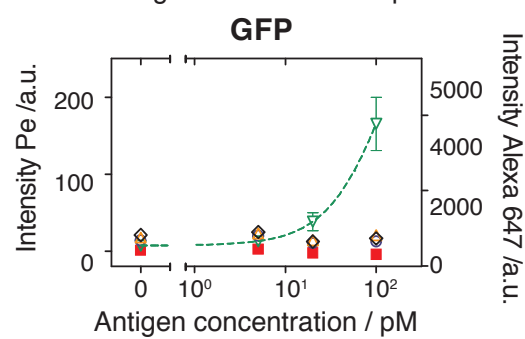

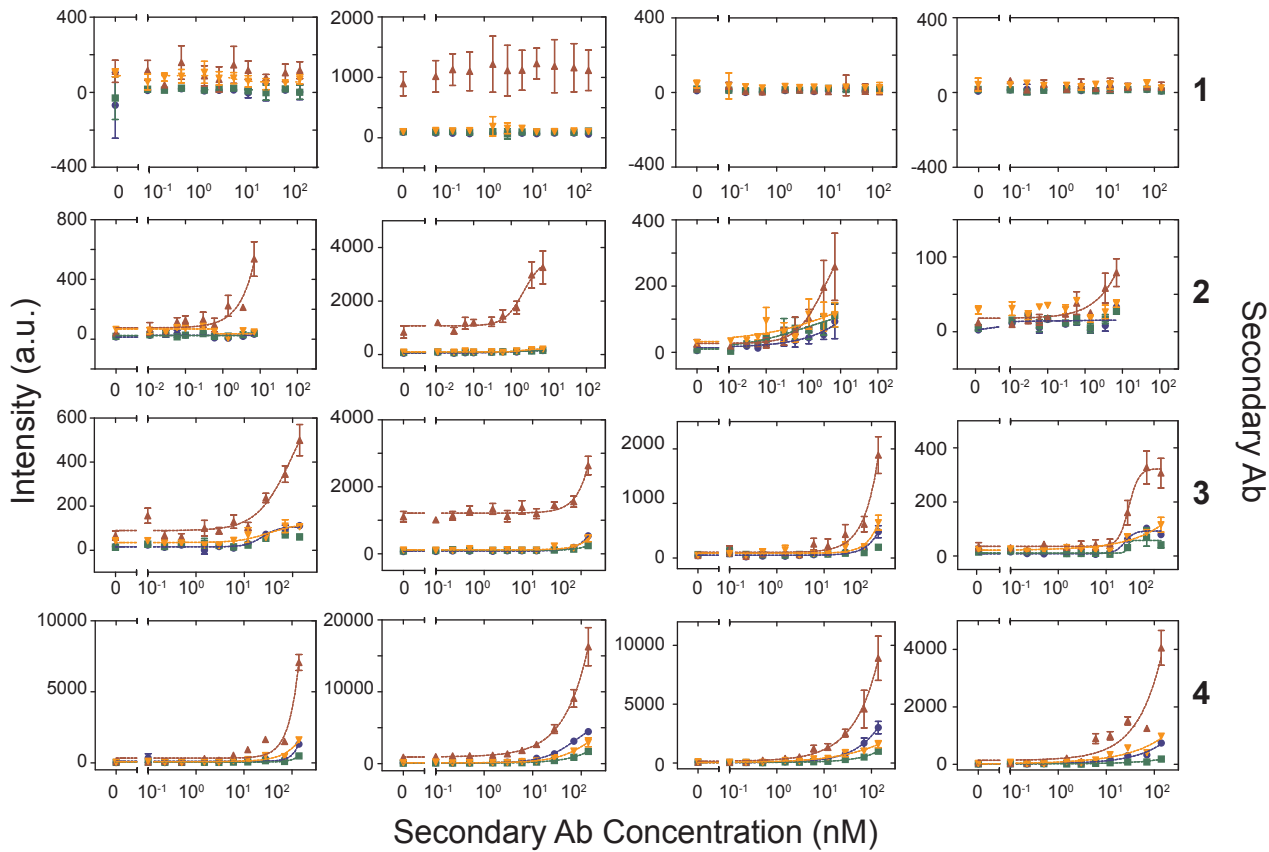

## 60nM Secondary Ab

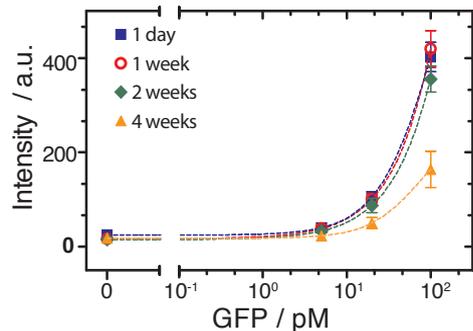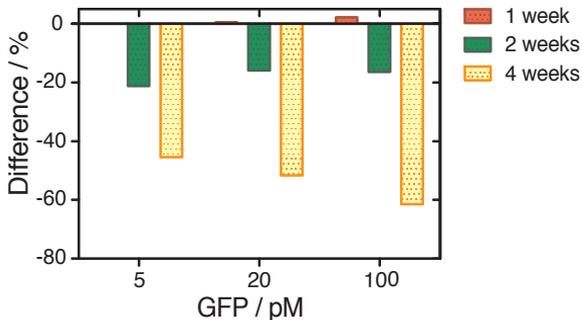

## 30nM Secondary Ab

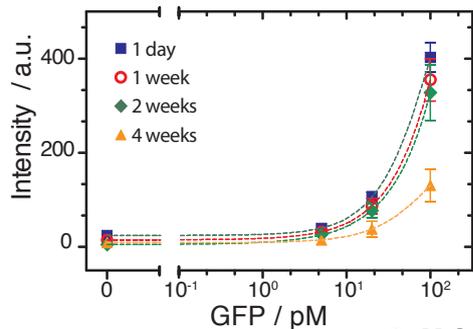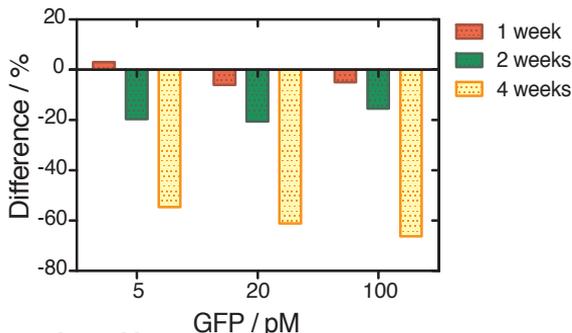

## 6nM Secondary Ab

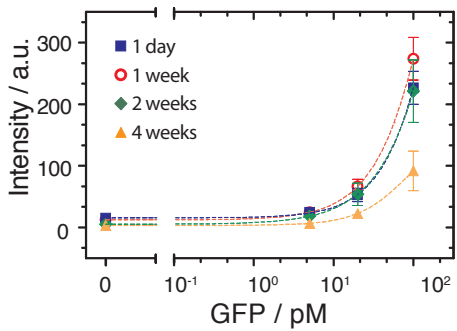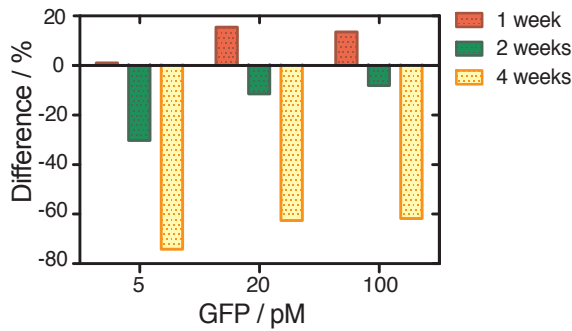

# Supplementary Information

# A microfluidic platform for high-throughput multiplexed protein quantitation

*Francesca Volpetti, Jose Garcia-Cordero, and Sebastian J. Maerkl*[*]

Institute of Bioengineering, School of Engineering, Ecole Polytechnique Federale de Lausanne, Lausanne, Switzerland.

E-mail: sebastian.maerkl@epfl.ch

**Figure S1 Fluorescent images of some of the unit cells on the device.**

All combinations of the five different primary antibodies and secondary antibodies were tested. A negative control (2% BSA in PBS) and a cocktail of proteins in buffer (GFP, IL-6, IL-1β, TNF-α, and PSA), at three different concentrations (5, 20, 100pM), were sequentially flowed. The color bars indicate the relative fluorescence intensity.

**Figure S1 Multiplex assay: Fluorescent signal and binding curves for 5 different proteins in solution: PSA, TNFα, IL-6, IL-1β and GFP as a function of different concentrations**.

For each graph the primary antibody was specific only for one antigen and the different curves are related to specific and non-specific secondary antibodies for the antigen. The secondary antibodies were spotted at a concentration of 2 nM. Left y-axis refers to secondary antibody labeled with phycoerythrin (Pe) and the right y-axis to secondary antibody labeled with Alexa Fluor ® 647 (error bars are std. dev., n=4-7).

**Figure S3 Screening of antibody combinations for PSA detection: secondary and tertiary antibodies were spotted.**

Different combinations of four secondary antibodies (1-4) and four tertiary antibodies (α– δ) were spotted. The secondary antibodies were spotted at six different concentrations, and the tertiary antibody at a concentration of 30 nM. After surface derivatization, four different primary antibodies (A-D) were flowed in sequence, one for each button, followed by 100 pM PSA in a buffer solution. Next the secondary antibody was allowed to diffuse in the reaction chamber and finally, the tertiary (α–δ) antibodies diffused in reaction chamber.

**Figure S4 Screening of antibody combinations for PSA detection: primary and secondary antibodies were spotted.**

16 different combinations of antibodies were spotted: four primary antibodies (A-B) and four secondary antibodies (1-4). The secondary antibodies were spotted at six different concentrations (0, 0.2, 1.2, 6, 28 and 140 nM), and the primary antibody at a concentration of 600 nM. After the surface derivatization, the primary antibody was allowed to diffuse and immobilize to the surface, then 100 pM PSA was flowed in the chip. Next the secondary antibody was allowed to diffuse in the reaction chamber and finally, three different tertiary antibodies (α, β, γ) were sequentially flowed, one for each button.

**Figure S5 Fluorescent signal intensity as a function of GFP concentration 1 day after generating the surface chemistry on the device (error bars are std. dev., n=9).**

The control lines were filled with oil (Fluorinert FC-40), and the functionalization of the surface was performed, as described. Next, the flow layer was dried by pushing air through the channel at 3.5 psi for 30 min. The chip was then stored in a vacuum chamber at room temperature for 1 day. To perform the immunoassay the chambers containing the spotted antibodies against GFP were filled with PBS in order to re-hydrate the spots. The unit cells were then isolated and the primary antibody allowed to diffuse into the reaction chamber. After a washing step, four different concentrations of GFP were flowed in sequence, one for each button. Finally, the secondary antibody was allowed to diffuse into the reaction chamber and the device was scanned.

**Table S1 List of antibodies.**

The table lists the 12 antibodies tested. Description, species, clonality, supplier and catalog number are provided for each antibody (Life-Technologies did not provide the clonality for antibodies α and β but they are likely polyclonal antibodies). The first set (A-D) includes the biotinylated primary antibodies, the second set (1-4) the secondary antibodies, and the third set (α−δ) the tertiary antibodies conjugated with phycoerythrin or Alexa Fluor 546.

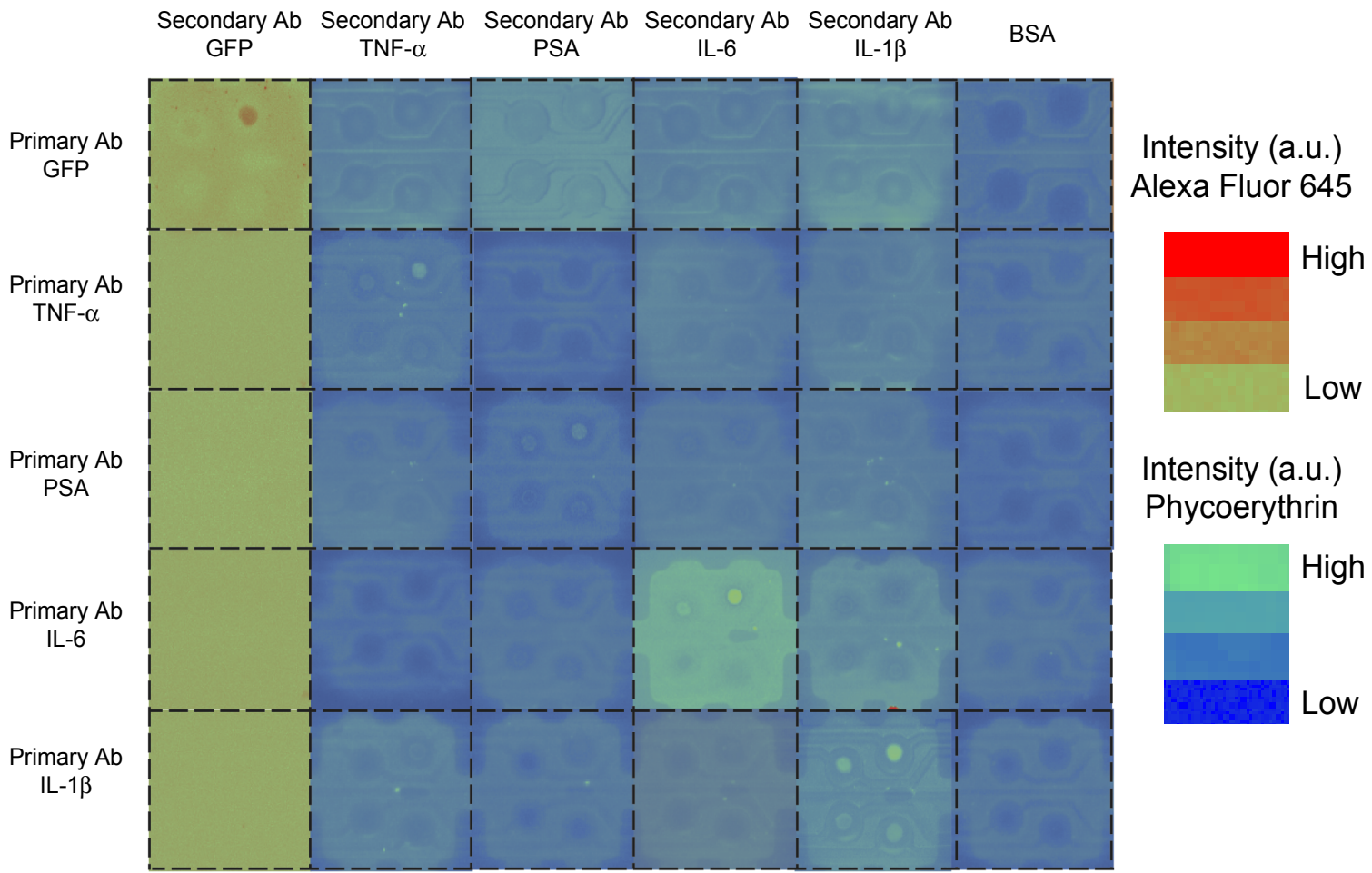

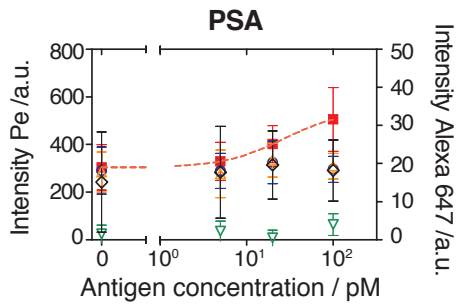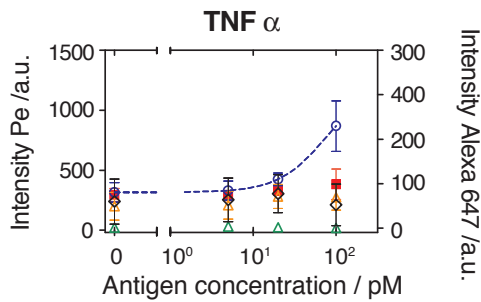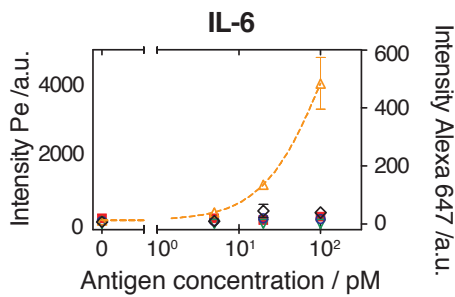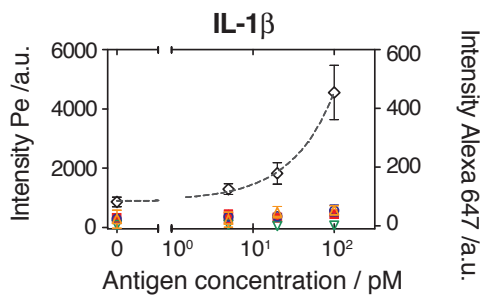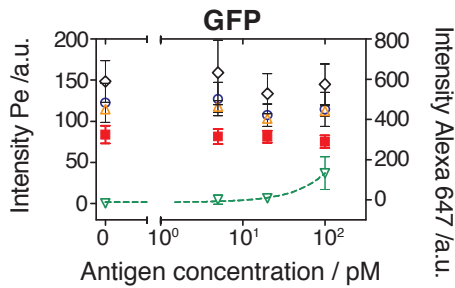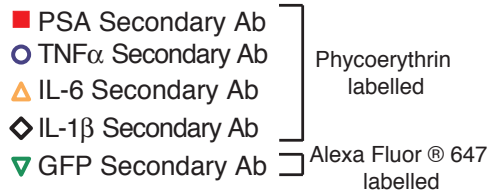

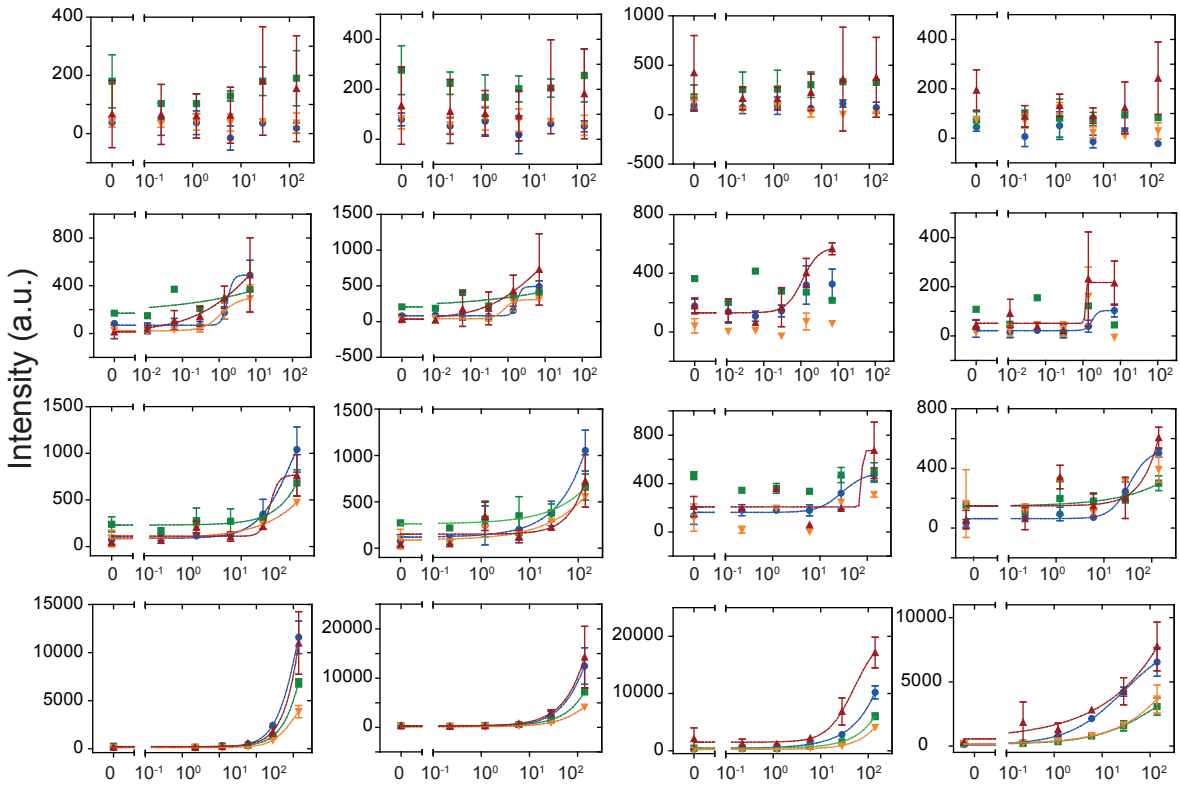

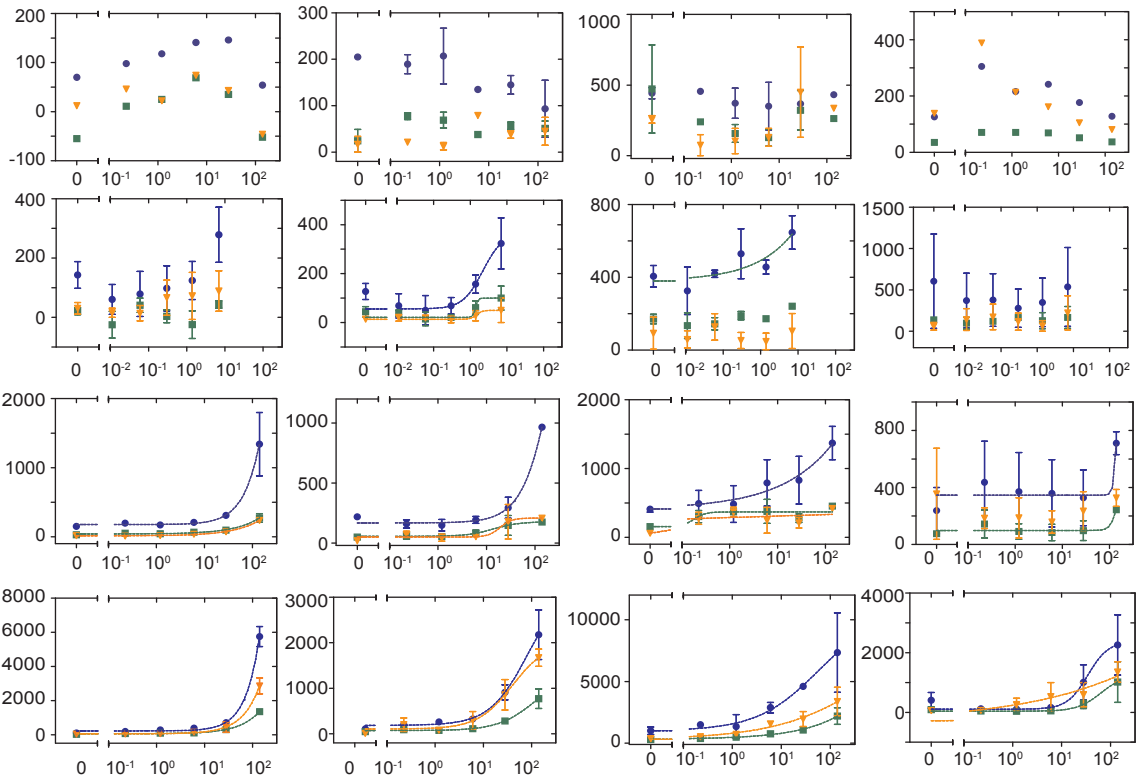

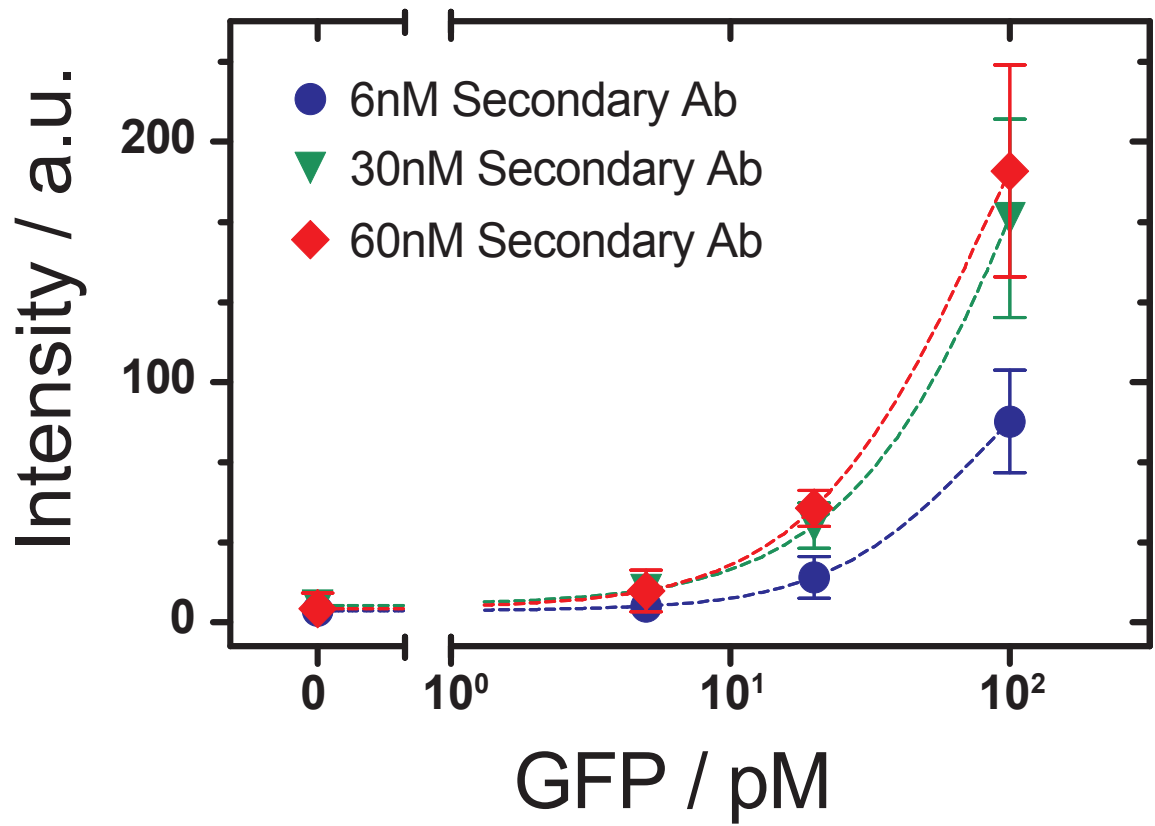

|   | Description | Species | Clonality | Supplier | Catalog Number |
|---|---|---|---|---|---|
| **A** | anti-PSA biotin conjugated | Mouse | monoclonal | Abcore | 1B-289 |
| **B** | anti-PSA biotin conjugated | Mouse | monoclonal | Abcam | ab77310 |
| **C** | anti-PSA biotin conjugated | Mouse | monoclonal | Abcam | ab182031 |
| **D** | anti-PSA biotin conjugated | Mouse | monoclonal | Abcam | ab182030 |
| **1** | anti-PSA | Rabbit | polyclonal | Fitzgerald | 70R-35481 |
| **2** | anti-PSA | Rabbit | monoclonal | Abcam | ab76113 |
| **3** | anti-PSA | Rabbit | polyclonal | Abcam | ab53774 |
| **4** | anti-PSA | Rabbit | polyclonal | Abcam | ab19554 |
| **α** | Goat anti-rabbit IgG PE conjugated | Goat | | Life Technologies | P-2771MP |
| **β** | Goat anti-rabbit IgG Alexa Fluor 546 conjugated | Goat | | Life Technologies | A-11010 |
| **γ** | Goat anti-rabbit IgG PE conjugated | Goat | polyclonal | Abcam | ab72465 |
| **δ** | Goat anti-rabbit IgG PE conjugated | Goat | polyclonal | Abcam | ab97070 |